\documentclass[usenatbib,12pt,letterpaper,notitlepage]{article}
\usepackage[margin=1in]{geometry}
\usepackage{graphicx}
\usepackage{amssymb}
\usepackage{amsmath}
\usepackage{amsbsy}
\usepackage{epstopdf}
\usepackage{pdflscape}
\usepackage{setspace}
\usepackage{nicefrac}
\usepackage{multirow}
\usepackage{booktabs}
\usepackage{natbib}
\usepackage{enumerate}
\usepackage{makeidx}  %%% standard INDEX
\usepackage[titletoc]{appendix}
\usepackage{fancyhdr}

\usepackage{longtable}

\usepackage[latin1]{inputenc}
\usepackage{geometry}
\usepackage{graphicx}
\usepackage{epstopdf}
\usepackage{amsmath}
\usepackage{amsfonts}
\usepackage{amssymb}
\usepackage{color}
\usepackage{bm}
\usepackage{float}
\usepackage{setspace}		% 1.5 spacing
\onehalfspace
\usepackage{hyperref}
\usepackage{subfigure}

\onehalfspace
\allowdisplaybreaks

\begin{document}

\title{Estimating Population Size with\\Link-Tracing Sampling}

\author{\begin{large}Kyle Vincent\end{large}\footnote{Currency Department, Bank of Canada, 234 Laurier Avenue West, Ottawa, Ontario, CANADA, K1A 0G9,\textit{email}: kvincent@bankofcanada.ca}
\begin{large}and Steve Thompson\end{large}\footnote{Department of Statistics and Actuarial Science, Simon Fraser University, 8888 University Drive, Burnaby, British Columbia, CANADA, V5A 1S6,\textit{email}: thompson@sfu.ca}
}
\date{\today}

\maketitle

\begin{abstract}
\noindent
We present a new design and inference method for estimating population size of a hidden population best reached through a link-tracing design.  The strategy involves the Rao-Blackwell Theorem applied to a sufficient statistic markedly different from the usual one that arises in sampling from a finite population.  An empirical application is described.  The result demonstrates that the strategy can efficiently incorporate adaptively selected members of the sample into the inference procedure.
\end{abstract}

\thanks{This work was supported through a Natural Sciences and Engineering Research Council Postgraduate Scholarship D and a Discovery Grant. The authors wish to thank Laura Cowen, Charmaine Dean, Maren Hansen, Chris Henry, Kim Huynh, Richard Lockhart, Louis-Paul Rivest, Carl Schwarz, and Jason Sutherland  for their helpful comments. The authors also wish to thank John Potterat and Steve Muth for making the Colorado Springs data available. All views expressed in this manuscript are solely those of the authors and should not be attributed to the Bank of Canada.}

%\thispagestyle{empty}

%\pagenumbering{gobble}

\pagenumbering{arabic}

\clearpage
\section{Introduction}

We introduce a new design-based method for estimating unknown quantities of hard-to-reach, networked populations when samples are selected through a link-tracing/adaptive sampling design. Since the population size is often unknown in hard-to-reach populations, we develop for such a situation a novel inference procedure based on a sufficiency result. In a typical sampling study, the usual minimal sufficient statistic for the population parameter vector is the unordered set of distinct units in the sample paired with their associated values of the variables of interest \citep{Thompson1996}. Yet for the current situation the standard sampling statistic is no longer sufficient.  We describe the new sufficient statistic and condition on it to obtain improved design-based estimators for the unknown population size.

Sampling from hard-to-reach populations, like those comprised of injection-drug users (IDUs), can be difficult and resource intensive as a large number of the individuals may be difficult to locate. Instead, recruitment can be based on tracing social links from members that have been selected for the sample to adaptively enlarge its size. Because these methods are practical for recruiting individuals in such settings, research for inferential methods based on adaptive sampling designs has found increasing acceptance in the literature; \cite{Thompson2006} and \cite{Handcock2010} outline design and model-based strategies, respectively, and \cite{Fienberg2010a} discusses papers with applications for sampling and analyzing hidden populations. However, hard-to-reach populations are typically not covered by a sample frame, rendering their size likely to be unknown. Consequently, many of these methods cannot be used to study the population.

Efficient inference for population size is an important factor in studying such populations, and hence link-tracing based strategies for making such inference have been growing in demand. However, most of these strategies developed for estimating population size are restricted to specific designs that do not permit much flexibility in adaptively selecting members. Furthermore, these methods are typically founded on model-based assumptions that complement the design so as to allow for ease in estimation of population size. As hidden populations will likely have a high degree of unpredictable behaviour (for example in the form of erratic clustering patterns among their members), model-based estimators may not be robust measures for the population size.

In contrast to the existing methods, our strategy has three primary advantages: (1) it grants the sampler the ability to choose how much sampling effort can be allocated towards conventional and adaptive selections; (2) it permits for flexibility in how members can be selected for the adaptive aspect of the sample selection procedure; and (3) it utilizes a design-based approach to inference to avoid dependence on model-based assumptions.

Design-based approaches have much potential to exploit the Rao-Blackwell theorem, a mathematically powerful technique that can be used to improve the precision of an estimator. The procedure entails exploiting a sufficient statistic to arrive at an improved estimator that retains the expectation of its preliminary counterpart while improving on its variance. The method outlined in this article consists of selecting independent adaptive samples and using standard estimation procedures, with the new sufficient statistic at the inference stage, to estimate population quantities like the size and mean. For a single-sample study, we make use of a design-based population size estimator presented in \cite{Frank1994} that parallels a mark-recapture approach in that it possesses a measure of overlap through counts of nominations originating from the initial sample. For a multi-sample study, we base the mark-recapture population size estimator on information in the randomly selected initial samples. Of note in both the single and multi-sample case this overlap may be small, which can make such estimates inefficient. Therefore, in our strategy we use the new sufficient statistic via the Rao-Blackwell method to weigh in the overlap among the traced parts of the sample(s). This method has the ability to preserve the bias while substantially increasing the efficiency of the estimators.

In Section 2 we introduce the notation used in this article, as well as outline and further explore a practical link-tracing sampling design (the Appendix provides details regarding the generalized sampling setup). In Section 3 we present the sufficiency result corresponding with the link-tracing sampling design outlined in Section 2 (the Appendix provides the corresponding sufficiency result for the generalized setup). Section 4 is reserved for developing estimators for the population size and mean, as well as those for the variances of these estimators. As tabulating the preliminary estimates from all reorderings of the final samples is computationally cumbersome for the samples selected in this study, in Section 5 we outline a Metropolis-within-Gibbs Markov chain resampling procedure to obtain approximations to the Rao-Blackwellized estimates. In Section 6 we perform a simulation study based on the empirical population. In Section 7 we draw conclusions and provide a general discussion of this novel method, including offering some ideas and direction for future work.

\section{Sampling Setup and Design}

Define $U=\{1,2,...,N\}$ to be the set of units/individuals that the population is comprised of, where $N$ is the population size. Define $y_i$ to be the response of interest of unit $i$. For example, in a drug-using population the response of interest could be an indicator variable based on the use of drug-injection equipment. In the network graph setting, each pair of units $(i,j)$, $i,j=1,2,...,N,$ is associated with a weight $w_{ij}$ which reflects the strength of the relationship from unit $i$ to unit $j$. For example, such a relationship could be based on the rate at which unit $i$ approaches unit $j$ to consume illegal drugs together through sharing drug-using equipment.

An adaptive sampling design which is selected without replacement typically consists of the selection of an initial sample and then further adaptive additions, and possibly conventional additions (for example, by taking random jumps; see the Appendix of this article for further information). In our study the design commences with the selection of an initial sample completely at random and is practical in that further recruitment is based only on tracing links. We outline the sample selection procedure in further detail below.

Suppose a study is based on $K$ samples. For each sample $k=1,2,...,K,$ where selection is based on an initial sample of size $n_{0k}$ and a desired final sample of size $n_k>n_{0k}$, the sample selection procedure is carried out as follows:

\bigskip
\noindent Step 0: Select $n_{0k}$ members completely at random.

\bigskip
\noindent Step $t$, $t=1,2,...,n_k-n_{0k}$: Define $s_{k,t}$ to be the set of currently sampled individuals for sample $k$ at step $t$. Let $a_{k,t}\subseteq s_{k,t}$ be the active set, namely, those individuals from whom we are considering tracing links, for sample $k$ at step $t$. Let $w_{a_{k,t},+}$ be the sum of the weights of the links from the active set to $U\setminus s_{k,t}$. If $w_{a_{k,t},+}=0$ (that is, there are no links out of the current sample) then the sampling procedure stops and the final sample is of size $n_{0k}+t-1$. If $w_{a_{k,t},+}>0$ then select an individual $i\ \epsilon\ U\setminus s_{k,t}$ with probability $q_{k,t,i}=\frac{w_{a_{k,t},i}}{w_{a_{k,t},+}}$ where $w_{a_{k,t},i}$ is the sum of the link weights from the active set out to unit $i$ at step $t$ for selection of sample $k$.

The observed data is $d_0=\{(i,y_i,w_{ij},w_i^+,t_{k,i}):i,j\ \epsilon\ s_k,\ k=1,2,...,K\}$ where $s_k$ refers to sample $k$ for $k=1,2,...,K$; $w_{ij}$ is the weight of the link from unit $i$ to unit $j$;  $w_i^+$ is the sum of the weights of all links emanating from individual $i$ (also known as the out-degree); and $t_{k,i}$ is the step in the sampling sequence when unit $i$ is selected for sample $k$. The probability of observing $d_0$ is expressed as
\begin{align}
p(D_0=d_0)=\prod\limits_{k=1}^K
\bigg(\frac{1}{{N\choose n_{0k}}}\prod\limits_{t=0}^{n_k-n_{0k}}q_{k,t,i}\bigg)
\end{align}

\noindent where the first term(s) in the expression corresponds with the random selection of the initial sample(s) and $q_{k,t,i}$ refers to the probability of selecting the unit selected for sample $k$ at step $t$. It shall be understood that for $t=0$ and $t > n_k-|s_k|$, $q_{k,t,i}=1$. Commencing the index with $t=0$ applies when only an initial sample is to be selected and no members are to be added adaptively to the corresponding sample.

\begin{figure}[H]
\begin{center}
\includegraphics [width=2.6 in,height=1.2 in]{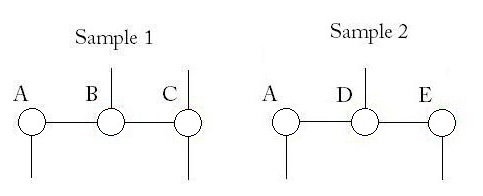}
\end{center}
\vspace{-5mm}
\caption{A two-sample study where samples are selected via the adaptive sampling design outlined in this section. The out-degree of each node is equal to the number of links emanating from the node. }
\label{jumpsexample}
\end{figure}

We clarify the sample selection procedure with the following illustration. Figure \ref{jumpsexample} provides an example of two final samples selected under the adaptive sampling design outlined in this section, where the study is comprised of two samples, thus $K=2$. The size of the initial samples are $n_{01}=n_{02}=1$ and the number of members added adaptively is two, to bring the final sample sizes up to $n_1=n_2=3$. In each case the active set is always the current sample.  For ease of understanding, we define $s_{(0_1,...,0_K)}$ to be the list of samples in the original order they are selected in.
\vspace{-5mm}

Suppose that links between nodes are reciprocated and the weight of each link is set equal to one. Further suppose that $s_{(0_1,0_2)}=((A,B,C),(A,D,E))$.  With a slight abuse of notation, we leave it implicit within the probability expressions that the required adjacency data is observed. The probability of selecting the samples in this order is
\begin{align}
p(s_{(0_1,0_2)})=\bigg(\frac{1}{N}\frac{1}{2}\frac{1}{3}\bigg)\times\bigg(\frac{1}{N}\frac{1}{2}\frac{1}{3}\bigg).
\end{align}

\section{Sufficiency Result}

Define $r$ to be the reduction function that maps the observed data to the reduced data $d_r$ via the removal of the time/step element assigned to each unit selected for each sample; $r(d_0)=d_r=\{(i,y_i,w_{ij},w_i+):i,j\ \epsilon\ s_k,\ k=1,2,...,K\}$. Hence, data reduction comes from mapping hypothetical observed data outcomes, in terms of reorderings of the sequence that the sampled members are selected in, to the reduced data corresponding with the original observed data. Below, we show that $d_r$ is a sufficient statistic for unobserved population quantities of the network; it is through averaging over estimates corresponding with reorderings that share mappings to the reduced data that one can obtain Rao-Blackwellized (improved) estimators of functions of such population quantities.

Index $x_k$ as $x_k=1,2,...,R_k$, $k=1,2,...,K$ where $R_k={|s_k| \choose n_{0k}}(|s_k|-n_{0k})!$  is the number of data reorderings under sample $k$. For each reordering $x_k$ of sample $k$ we define $q_{k,t,i}^{(x_k)}$ to be the probability of (hypothetically) adding that unit selected at step $t$ for sample $k$. We then define $s_{(x_1,...,x_K)}$ to be the list of the individually permuted samples in the order they are selected in.

\bigskip

\noindent \textbf{Theorem}: When samples are obtained with the sampling design outlined in the previous section, $D_r$ is a sufficient statistic for the population size, responses, and adjacency data.

\bigskip

\noindent \textbf{Proof}: Suppose $D_r=d_r$ is the reduced data. Choose any data reordering $s_{(x_1,x_2,...,x_K)}$.  The conditional probability of obtaining this data reordering is expressed as
\begin{align}
&p(s_{(x_1,x_2,...,x_K)}\mid d_r) = p(s_{(x_1,x_2,...,x_K)}) / \sum\limits_{r_1=1}^{R_1}\sum\limits_{r_2=1}^{R_2}\cdot\cdot\cdot\sum\limits_{r_K=1}^{R_K}p(s_{(r_1,r_2,...,r_K)})\notag \\
&=\frac{1}{{N \choose n_{01}}}\prod\limits_{t=0}^{n_1-n_{01}}q_{k,t,i}^{(x_1)}\times
\frac{1}{{N \choose n_{02}}}\prod\limits_{t=0}^{n_2-n_{02}}q_{k,t,i}^{(x_2)}\times\cdot\cdot\cdot\times
\frac{1}{{N \choose n_{0K}}}\prod\limits_{t=0}^{n_K-n_{0K}}q_{k,t,i}^{(x_K)} / \notag \\ &\sum\limits_{r_1=1}^{R_1}\sum\limits_{r_2=1}^{R_2}\cdot\cdot\cdot\sum\limits_{r_K=1}^{R_K}\bigg(
\frac{1}{{N \choose n_{01}}}\prod\limits_{t=0}^{n_1-n_{01}}q_{k,t,i}^{(r_1)}\times
\frac{1}{{N \choose n_{02}}}\prod\limits_{t=0}^{n_2-n_{02}}q_{k,t,i}^{(r_2)}\times\cdot\cdot\cdot\times
\frac{1}{{N \choose n_{0K}}}\prod\limits_{t=0}^{n_K-n_{0K}}q_{k,t,i}^{(r_K)}\bigg) \notag \\
&=\prod\limits_{t=0}^{n_1-n_{01}}q_{k,t,i}^{(x_1)}\times\prod\limits_{t=0}^{n_2-n_{02}}q_{k,t,i}^{(x_2)}
\times\cdot\cdot\cdot
\times\prod\limits_{t=0}^{n_K-n_{0K}}q_{k,t,i}^{(x_K)} / \notag \\
&\sum\limits_{r_1=1}^{R_1}\sum\limits_{r_2=1}^{R_2}\cdot\cdot\cdot\sum\limits_{r_K=1}^{R_K}
\bigg(\prod\limits_{t=0}^{n_1-n_{01}}q_{k,t,i}^{(r_1)}\times\prod\limits_{t=0}^{n_2-n_{02}}q_{k,t,i}^{(r_2)}
\times\cdot\cdot\cdot\times \prod\limits_{t=0}^{n_K-n_{0K}}q_{k,t,i}^{(r_K)}
\bigg).
\label{suffnojumps}
\end{align}

\noindent As this expression is independent of the population size, unobserved responses, and unobserved adjacency data, we can conclude that $D_r$ is a sufficient statistic for these quantities.

$\Box$

With respect to the example presented in Figure \ref{jumpsexample}, one pair of sample reorderings that is consistent with the sufficient statistic corresponding with the observed data is $s_{(x_1,x_2)}=((C,B,A),(D,A,E))$, for some pre-assigned $x_k=1,2,...,R_k$, where $R_k={3 \choose 1}(3-1)!=6$, $k=1,2$. Furthermore, the probability of selecting this reordering is
\begin{align}
p(s_{(x_1,x_2)})=\bigg(\frac{1}{N}\frac{1}{3}\frac{1}{4}\bigg)\times\bigg(\frac{1}{N}\frac{1}{3}\frac{1}{3}\bigg).
\end{align}

\noindent In contrast, one pair of sample reorderings that are not consistent with the sufficient statistic is $((C,A,B),(D,A,E))$, since it has zero probability of being selected due to an absence of a link to trace from unit $C$ to unit $A$ in the first sample.

\section{Estimation}

\subsection{Population size estimators}

Suppose that $\hat{N}_0$ is a preliminary estimate of the population size based on the original $K$ randomly selected initial samples (for example, see \cite{Frank1994} for a one-sample based approach in a network setting and \cite{Williams2002} for an overview of some commonly used multi-sample mark-recapture estimators). An improved estimator which has variance equal to or smaller than, and which shares the same expectation as, its preliminary counterpart is obtained via Rao-Blackwellizing the estimator over the sufficient statistic $d_r$. This estimator takes the form
\begin{align}
E[\hat{N}_0|d_r]=\hat{N}_{RB}=\sum\limits_{r_1=1}^{R_1}\sum\limits_{r_2=1}^{R_2}\cdot\cdot\cdot\sum\limits_{r_K=1}^{R_K}
\hat{N}_{0}^{(r_1,r_2,...,r_K)}p(s_{(r_1,r_2,...,r_K)}|d_r)
\end{align}

\noindent where $\hat{N}_{0}^{(r_1,r_2,...,r_K)}$ is the estimate of the population size based on the hypothetical initial samples corresponding with reorderings $r_1,r_2,...,r_K$ of samples $1,2,...,K$, respectively; $p(s_{(r_1,r_2,...,r_K)}|d_r)$ is the conditional probability of obtaining the sample reorderings $r_1,r_2,...,r_K$ given $d_r$.

\subsection{Population mean estimators}

Estimates of the distribution of individual responses, such as the proportion of injection-drug users or the average out-degree of the population members, are of interest to researchers of hard-to-reach populations; for example estimates for the rate of exchange of needles can be enhanced by such information \citep{Woodhouse1994}. We can obtain estimates of such population quantities as follows.  For notational convenience, we shall let $M=\bigcup\limits_{k=1}^K s_{0k}$. We can then estimate a population mean $y_{\mu}=\sum\limits_{i=1}^N y_i/N$ with the estimator based on the unique members selected for the initial samples, namely
\begin{align}
\hat{y}_0=\frac{\sum\limits_{i \epsilon M}y_i}{|M|}.
\label{preliminary mean}
\end{align}

\noindent Conditional on $|M|$ this estimator can be viewed as being based on a random sample of $|M|$ individuals selected without replacement. Therefore, $\hat{y}_0$ can be shown to be an unbiased estimator for $y_{\mu}$. The Rao-Blackwellized version of this estimator is obtained through the same procedure used to obtain that of the estimate of the population size; the corresponding formula for obtaining the Rao-Blackwellized version of $\hat{y}_0$ is, therefore,
\begin{align}
E[\hat{y}_0\mid d_r]=\hat{y}_{RB}=\sum\limits_{r_1=1}^{R_1}\sum\limits_{r_2=1}^{R_2}\cdot\cdot\cdot\sum\limits_{r_K=1}^{R_K}\hat{y}_0^{(r_1,r_2,...,r_K)}
p(s_{(r_1,r_2,...,r_K)}\mid d_r).
\end{align}

\subsection{Variance estimators}

\cite{Frank1994} outline several methods for obtaining estimators for the variance of the population size estimators they develop. Further, an abundance of literature exists on estimators for the variance of mark-recapture estimators; see \cite{Schwarz1999} and \cite{Amstrup2005} for such information. With respect to the population mean estimator, an estimate for the variance of $\hat{y}_0$ is the conditionally unbiased estimate
\begin{align}
\hat{\text{var}}(\hat{y}_0| |M|) = \bigg(\frac{N-|M|}{N}\bigg)\frac{s^2}{|M|},
\label{varyFPC}
\end{align}

\noindent where  $\frac{N-|M|}{N}$ corresponds with the finite population correction factor and \newline$s^2=\frac{1}{|M|-1}\sum\limits_{i\epsilon M} (y_i-\hat{y}_0)^2$. One caveat to using this approach is that the population size in Expression \eqref{varyFPC} must be replaced with a suitable estimate. In our empirical study we explore the use of mark-recapture estimators in lieu of the actual population size.

An unbiased estimate for the variance of a Rao-Blackwellized estimator can be obtained as follows. For any estimator $\hat{\theta}_{RB}=E[\hat{\theta}_0\mid d_r]$ for some population unknown $\theta$, where $\hat{\theta}_0$ is the preliminary estimate, the conditional decomposition of variances gives
\begin{align}
\text{var}(\hat{\theta}_{RB})= \text{var}(\hat{\theta_0}) - E[\text{var}(\hat{\theta}_0\mid d_r)].
\label{varRB}
\end{align}

\noindent An unbiased estimator for $\text{var}(\hat{\theta}_{RB})$ is
\begin{align}
\hat{\text{var}}(\hat{\theta}_{RB})=E[\hat{\text{var}}(\hat{\theta}_0)\mid d_r] - \text{var}(\hat{\theta}_0\mid d_r).
\label{varestRB}
\end{align}

\noindent This estimator is the difference of the expectation of the estimated variance of the preliminary estimator over all reorderings of the data and the variance of the preliminary estimator over all the reorderings of the data. As this estimator can result in negative estimates of the variance, a conservative approach is to take the estimate of $\text{var}(\hat{\theta}_{RB})$ to be $E[\hat{\text{var}}(\hat{\theta}_0)\mid d_r]$ on such occasions.

\section{Markov Chain Resampling Estimators}

When sample sizes are small, it would be ideal to enumerate all sets of sample reorderings with their corresponding preliminary estimates, as the improved estimators could then be obtained exactly. However, when sample sizes exceed computational feasibility for exact enumeration, a Markov chain Monte Carlo (MCMC) resampling procedure %(\cite{Metropolis1953} and \cite{Hastings1970})
can be implemented to obtain approximations to the Rao-Blackwellized estimates. We outline such a procedure below, namely a Metropolis-within-Gibbs accept/reject resampling procedure. The strategy entails considering a candidate reordering of only one sample at each step of an iteration in order to encourage mixing in the chain.

\bigskip

\noindent Suppose $\theta$ is an unknown population quantity we wish to estimate with the improved estimator $\hat{\theta}_{RB}=E[\hat{\theta}_0\mid d_r]$ where $d_r$ is a sufficient statistic for $\theta$.

\bigskip

\noindent Step 0: The MCMC procedure commences in its stationary distribution, with estimates based on the original observed data, so that it will remain in its stationary distribution at each iteration; let $\hat{\theta}_{0}^{(0,...,0)}$ be the estimated value of $\theta$ and $\hat{\text{var}}(\hat{\theta}_{0}^{(0,...,0)})$ be the estimated value of $\text{var}(\hat{\theta}_0)$ obtained from the $K$ adaptive samples in the original order they were selected. Note that there are $K$ copies of the zeros in the exponents. Let $t_k^{(0)}=s_{(0_k)}$, that is, the original sample in the order it was selected, for $k=1,...,K$.

\bigskip

\noindent For step $l = 1,2,...,R$, where $R$ is sufficiently large, and $k=1,...,K$: Draw a candidate reordering of sample $k$, $t_k^*$ say, from a candidate distribution $q_k$ that corresponds with sample $k$. Suppose the most recently accepted candidate reordering of sample $k$ is $t_k^{(z_k)}$ for some reordering of the sample where $z_k=0,1,2,...,l-1$. Let $p(t_k^*)$ be the empirical probability of obtaining $t_k^*$ and let $q_k(t_k^*)$ be the probability of obtaining $t_k^*$ under sample $k$'s corresponding candidate distribution. With probability equal to $\text{min}\bigg\{\frac{p(t_k^*)}{p(t_k^{(z_k)})}\frac{q_k(t_k^{(z_k)})}{q_k(t_k^*)},1\bigg\}$
\noindent let $\hat{\theta}_0^{(l,...,l,l,l-1,...,l-1)}$ and $\hat{\text{var}}(\hat{\theta}_{0}^{(l,...,l,l,l-1,...,l-1)})$ be the estimates of $\theta$ and $\text{var}(\hat{\theta}_0)$, respectively, obtained with the ordered set of sample reorderings $(t_1^{(l)},...,t_{k-1}^{(l)},t_{k}^{*},t_{k+1}^{(l-1)}...,t_K^{(l-1)})$, and set $t_k^{(l)}=t_k^*$. Otherwise, take $\hat{\theta}_{0}^{(l,...,l,l,l-1,...,l-1)}=\hat{\theta}_{0}^{(l,...,l,l-1,l-1,...,l-1)}$, \newline$\hat{\text{var}}(\hat{\theta}_{0}^{(l,...,l,l,l-1,...,l-1)})=\hat{\text{var}}(\hat{\theta}_{0}^{(l,...,l,l-1,l-1,...,l-1)})$, and set $t_k^{(l)}=t_k^{(l-1)}$. Recall that with the adaptive sampling design outlined in this paper, $p(t_k^{(l)})$ need only be evaluated up to the (hypothetical) adaptive recruitment probabilities corresponding with the ordered set of sample reorderings. The reason is that the terms involving the unknown population size $N$ can be factored out of the ratio of the empirical probabilities of obtaining any ordered sample reorderings and canceled from the expression; see Expression \eqref{suffnojumps} for further details.

\bigskip

\noindent Final step: An enumerative estimate of $\hat{\theta}_{RB}$ is
\begin{align}
\tilde{\theta}_{RB}&=\frac{1}{R}\sum\limits_{l=1}^{R}\frac{1}{K}\sum\limits_{k=1}^K \hat{\theta}_{0}^{(l,..,l,l,l-1,...,l-1)},
\end{align}
\noindent where the last $l$ in the exponent refers to the $k^{\text{th}}$ entry in the vector, which corresponds with the index value. Similarly, an enumerative estimate of $\hat{\text{var}}(\hat{\theta}_{RB})$ is
\begin{align}
\tilde{\text{var}}(\hat{\theta}_{RB})&
=\tilde{E}[\hat{\text{var}}(\hat{\theta}_0)\mid d_r]-\tilde{\text{var}}(\hat{\theta}_0\mid d_r)\notag \\
&=\frac{1}{R}\sum\limits_{l=1}^R\frac{1}{K}\sum\limits_{k=1}^K\hat{\text{var}}(\hat{\theta}_{0}^{(l,..,l,l,l-1,...,l-1)})-\frac{1}{R}\sum\limits_{l=1}^R
\frac{1}{K}\sum\limits_{k=1}^K (\hat{\theta}_{0}^{(l,..,l,l,l-1,...,l-1)}-\tilde{\theta}_{RB})^2.
\end{align}

\bigskip

In our study we explore the use of a candidate selection distribution for all sample reordering selections that mimics the sample selection procedure outlined in Section 2, to encourage efficiency in the resampling estimation procedure, as follows. First, place all sampled units not nominated by any other sampled units into the hypothetical initial sample with probability one. Note that these members must be in the corresponding original initial sample, otherwise they cannot be selected for the sample under this design. Next, fill in the rest of the initial sample completely at random with members not yet selected. Finally, attempt to select the remaining members using the same design that gave rise to the original sample.

In the event that the final sample size is less than that which was pre-specified, that is, $|s_k|<n_k$, this will be due to an absence of links from the active set out of the final sample. Careful attention must then be paid to determining which reorderings are consistent with the sufficient statistic since such sample reorderings must have a final active set that does not reach out of the final sample. In the event a reordering has additional links out of its final active set, continuing with sampling would be permitted. This in turn will result in a larger, and therefore different, final sample than that originally obtained, and such a reordering would not be consistent with the sufficient statistic.

\section{Empirical Study}

\begin{figure}[H]
  \begin{center}
    \mbox{
          \subfigure{\includegraphics[width=6in]{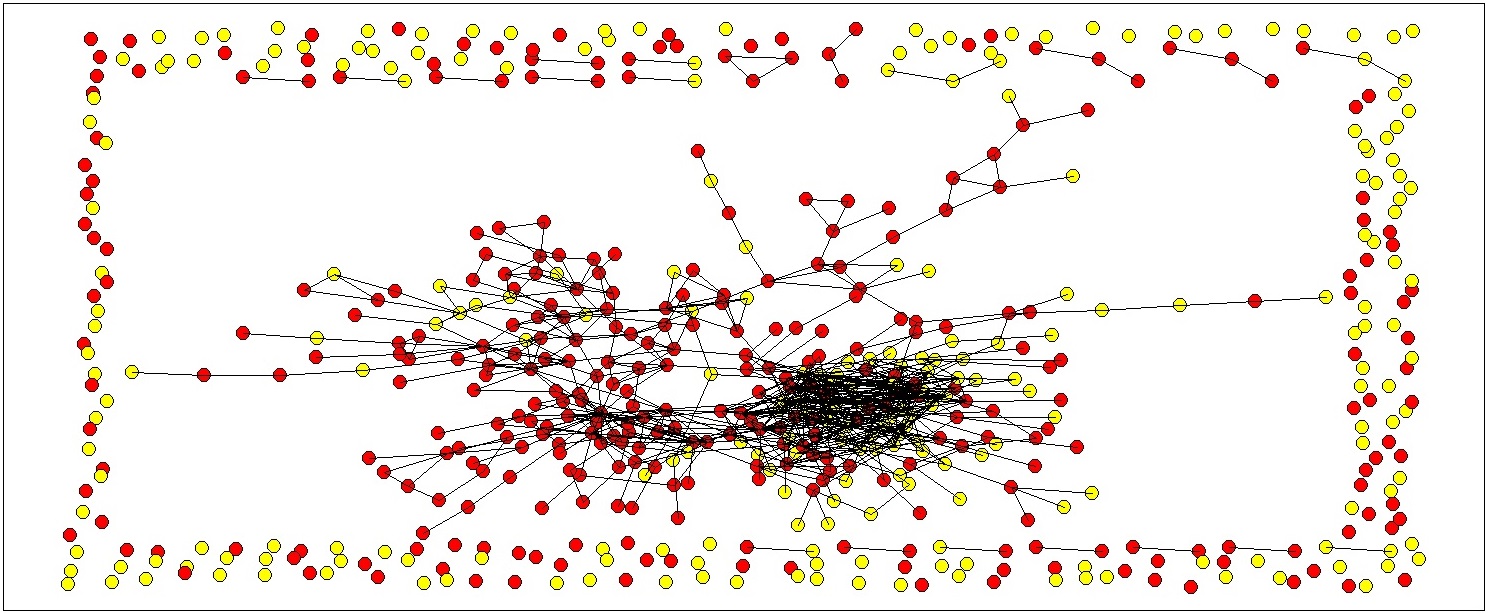}}\quad
          }
          \vspace{-5mm}
    \caption{HIV/AIDS at-risk population \citep{Potterat1993}. The dark-coloured nodes indicate individuals who are injection drug-users and the light-coloured ones indicate non-injection drug-users. Links between pairs of nodes indicate drug-using relationships and links are reciprocated. The size of the population is 595, the proportion of injection drug-users is 0.575, and the average out-degree is 2.45.}
    \label{fig12a}
  \end{center}
\end{figure}

To evaluate the link-tracing sampling design and new inference procedure outlined in this article, we use a simulation study on an empirical population of individuals at risk for HIV/AIDS in the Colorado Springs area \citep{Darrow1999, Potterat1993, Rothenberg1995}. The population data is based on Project 90, a prospective study funded by the Center for Disease Control and Prevention, and is summarized in Figure \ref{fig12a}; for more details on Project 90, see https://opr.princeton.edu/archive/p90/.
\vspace{-5mm}
\vspace{-5mm}
Figure \ref{fig12} shows two adaptive samples, each independently obtained with the design outlined in Section 2. Notice the additional, and disproportionate, overlap in the final samples that can be exploited for inferential purposes with the strategy presented in this paper.
\vspace{-5mm}
\begin{figure}[H]
  \centering
    \mbox{
      \subfigure{\includegraphics[width=5.7in]{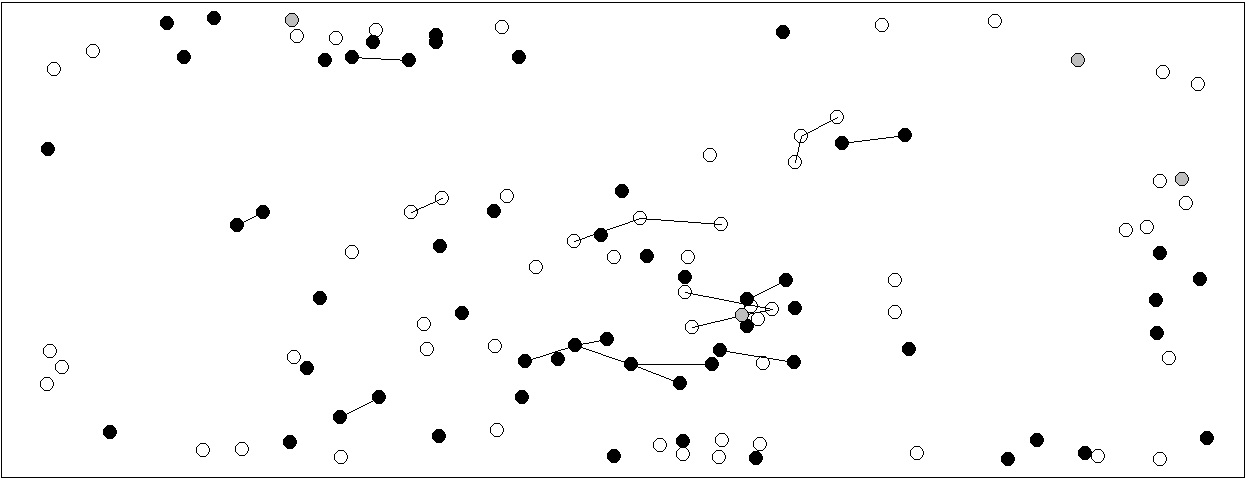}}}
\end{figure}
\vspace{-5mm}
\begin{figure}[H]
  \centering
     \mbox{
      \subfigure{\includegraphics[width=5.7in]{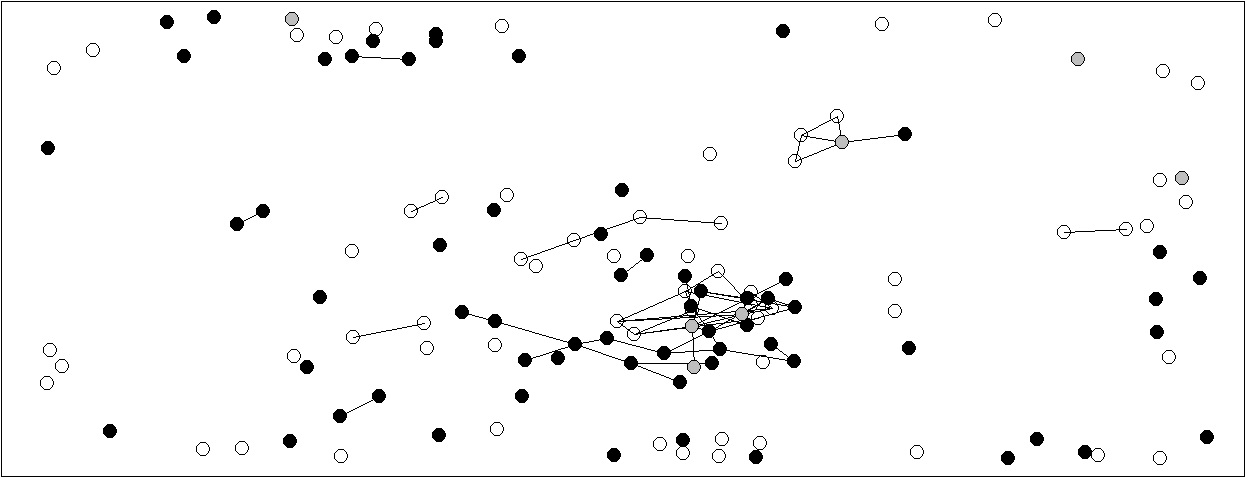}}
      }
      \vspace{-5mm}
\caption{Top: Two random initial samples, each of size 60. Bottom: Two final adaptive samples with 10 members added adaptively. Links between individuals are reciprocated and equal to one. All individuals are defined to have no link to themselves. The active set is always the current sample. Samples are distinguished by the colours of the nodes, with grey nodes representing individuals selected for both samples. Four individuals are selected for both initial samples, and seven are selected for both final samples. The preliminary bias-adjusted Lincoln-Petersen estimator \citep{Chapman1951} for the population size is $\hat{N}_0=743$ and its improved counterpart is $\hat{N}_{RB}=671$.
}
    \label{fig12}
\end{figure}
%I SET THE SEED EQUAL TO 6 TO GET THESE PLOTS (AND SCORES)

A one-sample, two-sample, and three-sample simulation study is conducted based on the following parameters. We set $w_{ij}= w_{ji} =1$ in the presence of a link between units $i$ and $j$, and zero otherwise. We also define $w_{ii}=0$ for all $i=1,2,...,N$. We define $y_i=1$ if individual $i$ is an injection-drug user, and zero otherwise. We explore the use of the sampling design outlined in Section 2 for the independent selection of each sample. In each study, 2500 simulation runs are obtained where initial samples are of size 60 and (up to) 10 members are recruited adaptively. In each study the active set is always the current sample. Approximations to the improved estimators are based on the MCMC method outlined in the previous section. The sample reordering proposals are found to be accepted at a rate of approximately 10\%. We therefore base inference on 10,000 resamples corresponding with each sample (thus, we set the value of $R$ defined in the previous section equal to $10,000$).

In the one-sample study we utilize a design-based population size estimator developed by \cite{Frank1994}, namely that which they denote as $\hat{\nu}_5$. This estimator is a function of the initial sample size, number of links within the initial sample (which is analogous to the recaptures in a mark-recapture study), and number of links from the initial sample to outside the initial sample. When selecting an adaptive sample in a network setting where the weight of each link is one, we can see that the statistics required for this estimator conform with our setup and hence the estimator can be Rao-Blackwellized under the sufficient statistic we outline. In our study, we mimic the approach used by \cite{Chapman1951} to adjust the Lincoln-Petersen estimator by adding a value of one to each aforementioned statistic in order to stabilize the estimator. We base an estimator for the variance of this estimator on the same jackknife routine that the authors use for their aforementioned estimator.

In the two-sample study we utilize the bias-adjusted Lincoln-Petersen population size estimator \citep{Chapman1951}. We take as an estimator for the variance of this estimator that which is proposed by \cite{Seber1970}.

In the three-sample study we make use of the population size estimators, and corresponding estimators of the variance of these estimators, provided by the `Rcapture' package \citep{Rivest2012}. In particular we use those estimates bias-corrected through frequency modifications (see \citealp{Rivest2001} for further details), namely: the maximum likelihood estimator based on a log-linear model \citep{Baillargeon2007}; the \cite{Chao1987} $M_h$ lower bound estimator; the Poisson2 (using a Poisson model) estimator based on an $M_h$ assumption \citep{Baillargeon2007}; Darroch's $M_h$  estimator \citep{Darroch1993}; and the Gamma3.5 (using a Gamma model) estimator based on an $M_h$ assumption \citep{Baillargeon2007}.

In each study we explore the estimator for the population proportion of injection drug-users, as well as the average out-degree by replacing $y_i$ with $w_i^+$ in Expression \eqref{preliminary mean}. We take as estimators for the variance of these estimators that proposed in Expression \eqref{varyFPC}.

The one-sample design-based estimator, bias-adjusted Lincoln-Petersen estimator, and maximum likelihood estimator are all based on a homogenous sampling model, namely the $M_0$ model.  Each of the other population size estimators is based on a heterogenous sampling model, the $M_h$ model, which rests on the assumption that selection probabilities differ between individuals within each sample (see \cite{Chao2001} for more details). Though our study is based on homogenous selection probabilities, we shall explore the use of the estimators based on the $M_h$ model to gauge the increase in precision of the improved estimators.

Table \ref{estimate23} provides the approximate expectation and variance of the preliminary and improved estimators, as well as the approximate variance scores based on a random sample of size 70. In all cases a significant improvement is seen with the Rao-Blackwellized estimators relative to their preliminary counterparts. Furthermore, the improved estimates offer a competitive alternative to the use of estimators based on random samples of size moderately larger than the initial sample sizes.

\begin{table}[H]
\centering
\caption{Approximate expectation and variance of preliminary and improved estimators, and variance of estimators based on random samples of size equal to the desired final sample size. The size of each initial sample is 60 and the desired final sample sizes are 70. Entry ``Proportion of IDUs" refers to the unbiased estimate for the proportion of individuals in the population who are injection drug-users. Entry ``Average node-degree" refers to the unbiased estimate for the average out-degree. All other entries refer to estimators for the population size of 595.}
\begin{tabular}{l*{6}{r}r}
Estimator                       &Expectation        &Var., Prelim.              &Var., Improved             &Var., RS\\\hline
One-sample study:\\\hline
Frank and Snijders' estimator   &705                &172925                     &102190                     &69672\\
Proportion of IDUs              &0.575              &0.00368                    &0.00334                    &0.00326\\
Average node-degree             &2.45               &0.23802                    &0.18102                    &0.23509\\
\hline
Two-sample study:\\\hline
Lincoln-Petersen                &592                &60162                      &49061                      &40919\\
Proportion of IDUs              &0.575              &0.00176                    &0.00158                    &0.00145\\
Average node-degree             &2.45               &0.11314                    &0.08018                    &0.09765\\
\hline
Three-sample study:\\\hline
Maximum likelihood $M_0$        &593                &15843                      &13021                      &11058\\
Chao LB                         &592                &18372                      &14668                      &12459\\
Poisson2                        &622                &169719                     &113609                     &87949\\
Darroch                         &603                &1277823                    &649095                     &374411\\
Gamma3.5                        &702                &9534438                    &3314293                    &1194259\\
Proportion of IDUs              &0.575              &0.00109                    &0.00098                    &0.00091\\
Average node-degree             &2.45               &0.07218                    &0.04985                    &0.06025\\
\hline
\label{estimate23}
\end{tabular}
\end{table}

Table \ref{coverage23} provides the coverage rates for the population size, proportion of injection drug-users, and average out-degree corresponding with the aforementioned estimators when using nominal 95\% confidence intervals based on the Central Limit Theorem (CLT), as well as a log transformation strategy outlined in Chao (1987) for population size estimators. With respect to the estimators for the variance of the estimators of the population proportion and average out-degree, we substitute Frank and Snijders' estimator from the one-sample study, the bias-adjusted Lincoln-Petersen estimator from the two-sample study, and the Maximum likelihood $M_0$ estimator from the three-sample study into the corresponding variance expression found in Expression \eqref{varyFPC}. The high coverage rates of these estimators indicate this is a suitable choice. A small number of negative estimates for the variance of the improved estimates are found with Frank and Snijders' estimator. A moderate number of negative estimates for the variance of the improved estimates are found with the Darroch and Gamma3.5 estimators; these estimators are rather unstable, as reflected upon by their variance scores when compared to the other estimators. Consequently, in resorting to using the conservative approach suggested in Section 4 (see the discussion after Expression \eqref{varestRB}), we find that coverage rates for these estimators are higher than those based on their preliminary counterparts. In all other cases the coverage rates of the confidence intervals based on the improved estimators are on par with their preliminary counterparts.

\begin{table}[H]
\centering
\caption{Coverage rates of confidence intervals corresponding with estimators of the preliminary and improved estimators, with average length of intervals in parentheses. Entry ``Proportion of IDUs" refers to the unbiased estimate for the proportion of individuals in the population who are injection drug-users. Entry ``Average node-degree" refers to the unbiased estimate for the average out-degree. All other entries refer to estimators for the population size of 595.}
\begin{tabular}{l*{6}{l}l}
Estimator                       &CLT Prelim.     &CLT Improved   &Log Prelim.     &Log Improved \\\hline
One-sample study:\\\hline
Frank and Snijders' estimator   &0.934 (1634)    &0.970 (1822)   &0.970 (1996)    &0.981 (2229) \\
Proportion of IDUs              &0.945 (0.237)   &0.946 (0.227)  &                &     \\
Average node-degree             &0.912 (1.870)   &0.935 (1.643)  &                &     \\
\hline
Two-sample study:\\\hline
Lincoln-Petersen                &0.860 (800)     &0.875 (742)    &0.922 (854)     &0.920 (785)\\
Proportion of IDUs              &0.938 (0.161)   &0.942 (0.153)  &                &     \\
Average node-degree             &0.927 (1.299)   &0.940 (1.105)  &                &     \\
\hline
Three-sample study:\\\hline
Maximum likelihood $M_0$        &0.918 (483)    &0.925 (445)     &0.929 (498)     &0.945 (456)  \\
Chao LB                         &0.909 (494)    &0.911 (449)     &0.933 (511)     &0.930 (461)  \\
Poisson2                        &0.919 (1485)   &0.943 (1413)    &0.972 (1802)    &0.982 (1691) \\
Darroch                         &0.453 (2073)   &0.595 (2266)    &0.437 (2736)    &0.663 (3095) \\
Gamma3.5                        &0.439 (3822)   &0.617 (4661)    &0.436 (5599)    &0.678 (7277) \\
Proportion of IDUs              &0.947 (0.129)  &0.947 (0.121)   &                & \\
Average node-degree             &0.930 (1.039)  &0.944 (0.873)   &                & \\
\hline
\label{coverage23}
\end{tabular}
\end{table}

\section{Discussion}

In this article we outline a new strategy which uses link-tracing sampling and design-based inference to estimate the population size and other important characteristics of networked, hard-to-reach populations. The new method possesses the ability to adaptively recruit individuals for the study without introducing additional bias into the inference procedure, while allowing for control over sample sizes. As the theoretical results and simulation studies show, the novel method outlined here gives rise to more precise estimators relative to those based on the random initial samples.

The adaptive sampling design outlined in Section 2 bases recruitment after selecting the initial sample entirely on tracing links. However, adaptive sampling designs are not necessarily restricted to tracing links for further recruitment. In many cases it would be advantageous to allow for random jumps to be taken at intermediate steps of the selection process, perhaps when wishing to avoid over-sampling in heavily connected components of the network and/or to allow sampling to continue when links out of the active set are exhausted. Such a general case of adaptive sampling designs, with a sufficiency result, are detailed in the Appendix. As we show, the sufficient statistic for this case reflects when random jumps are made, a direct implication of the population size being unknown. In contrast to the minimal sufficiency result in sampling when the population size is known (see \cite{Thompson1996} for further details), the sufficiency result now depends on the adaptive sampling design that is implemented; we can compare the sufficient statistic outlined in Section 3 with that in the Appendix. Furthermore, the theory outlined in this paper reveals an interesting result. In the usual survey sampling setting when the population size is known and sampling is based on an adaptive design, the likelihood function of the unknown responses is flat \citep{Thompson1996}. Yet in the case when the population size is unknown and sampling is based on an adaptive design, Expressions \eqref{suffnojumps} and \eqref{likelihoodjumps} in the Appendix show that the likelihood is not flat.

An advantage this new method possesses over existing methods for population size estimation is outlined as follows. In some empirical settings when sample sizes are small, the selection of random samples may give rise to little or no overlap, or nominations within the initial sample in the case of a single-sample study, rendering an inflated and therefore undesirable estimate of the population size when using a mark-recapture style of estimator. With the method outlined in this article, overlap in the adaptive recruitment stage of the sample selection procedure is more certain and hence the use of the new inferential procedure should result in a more reliable and stable estimate of the population size.

Expression \eqref{varRB} reveals that the greater the variability amongst the preliminary estimates corresponding with sample reorderings, the greater the expected improvement in the Rao-Blackwellized estimators. Hence, in some situations it may be advantageous to steer the sampling design, possibly through a choice of active set and/or parameter(s) corresponding with random jumps, in order to encourage reorderings that both vary amongst their corresponding preliminary estimates and are consistent with the sufficient statistic. One possible route to explore is that which restricts the active set to members that identify a relatively large number of peers, since there is likely to be more heterogeneity in the measured overlap amongst reorderings, as well as more consistent reorderings. Further research on this topic would be highly beneficial.

%Further research on suitable candidate distributions for the MCMC procedure used to approximate the improved estimators will be required. With respect to the candidate distribution we outlined, as the size of the study grows (in terms of the number of samples) the probability a proposal will be accepted will likely decrease since the sample path also grows. One possible direction for future exploration would be to allow for the chain to take a random jump to a different consistent sample reordering when it stagnates for a large number of proposals. Such an approach would likely encourage mixing in the chain and hence would require less computational effort to obtain reliable approximations to the improved estimators.

%In this article we have considered a design-based Rao-Blackwellization approach to inference. We highlight here that the inference strategy requires observing all the adjacency data within the final sample. In many empirical settings it may be difficult to measure exactly how many peers of a selected unit are reachable and, perhaps due to privacy concerns, even more difficult to determine which members of the sample identify each other. Future work on strategies that make use of model-assisted design-based approaches \citep{Sarndal2003} when it is anticipated that such stringent assumptions cannot be met is deserving of attention. Such a contribution has already been made in the literature on respondent-driven sampling by \cite{Gile2011}.

In the empirical setting it is likely that the selection procedure for the initial samples would not be completely random. For example, there may be a propensity for self-selection amongst individuals for a study. Hence, it may help to introduce an element of heterogeneity in sampling to account for this via selection probabilities that are heterogenous between strata (see \cite{Chao2001} for further information on the $M_h$ mark-recapture model). In this case we may assume there are $G$ strata that the population are divided into, and thus choose a predetermined number of individuals, $n_{0k,g}$ say, to be selected for initial sample $k=1,2,...,K$ from stratum $g$, $g=1,2,...,G$. When samples are selected only by tracing links after the initial samples are selected, the original data is then $d_0=\{(i,g_i,y_i,w_{ij},w_i^+,t_{k,i}):i,j\ \epsilon\ s_k,\ k=1,2,...,K\}$ where $g_i$ is the stratum unit $i$ belongs to. Keeping with the design-based approach to inference it can then be shown that $d_r=\{(i,g_i,y_i,w_{ij},w_i+):i,j\ \epsilon\ s_k,\ k=1,2,...,K\}$ is a sufficient statistic for the population size, unobserved responses, and unobserved adjacency data. In this case, reorderings  consistent with the sufficient statistic must have $n_{0k,g}$ units selected for the initial component of sample $k$. Note that this is left implicit in the sampling design and therefore does not need to be reflected upon in the observed, and hence reduced, data. Extending the methods outlined in this article in a similar fashion to work with more elaborate closed population mark-recapture models \citep{Schwarz1999} will make for interesting future work.

%%%%%%  bibliography

%\newpage
\bibliographystyle{biom}
%\addcontentsline{Bibliography}
\bibliography{MasterReferences}

\appendix
\section{Appendix}

This Appendix outlines the adaptive sampling design that permits for random jumps at intermediate stages of the sample selection procedure. It also presents a sufficiency result for this design and provides a discussion of sample reorderings that are consistent with this sufficient statistic.

\subsection{Adaptive sampling design that allows for random jumps}

Suppose a study is based on $K$ samples. For each sample $k=1,2,...,K,$ where selection is based on an initial sample of size $n_{0k}$ and a final sample of size $n_k>n_{0k}$, the sample selection procedure is carried out as follows:

\bigskip
\noindent Step 0: Select $n_{0k}$ members completely at random.

\bigskip
\noindent Step $t$, $t=1,2,...,n_k-n_{0k}$: Define $s_{k,t}$ to be the current sample and let $a_{k,t}\subseteq s_{k,t}$ be the active set at step $t$. Let $w_{a_{k,t},+}$ be the sum of the weights of the links from the active set to $U\setminus s_{k,t}$. If $w_{a_{k,t},+}=0$, that is, there are no links out of the current sample, then select a unit $i\ \epsilon\ U\setminus s_{k,t}$ with probability $\frac{1}{N-(n_{0k}+t-1)}$, and thus a random jump is forced at this step. If $w_{a_{k,t},+}>0$ then with probability $d$ select a unit $i\ \epsilon\ U\setminus s_{k,t}$ with probability $q_{k,t,i}=\frac{w_{a_{k,t,i}}}{w_{a_{k,t},+}}$, and with probability $1-d$ select a unit $i\ \epsilon\ U\setminus s_{k,t}$ with probability $\frac{1}{N-(n_{0k}+t-1)}$; in the latter case, again, a random jump is taken.

The observed data is $d_0=\{(i,y_i,w_{ij},w_i^+,t_{k,i}),\underline{J}_k,\underline{H}_k:i,j\ \epsilon\ s_k,\ k=1,2,...,K\}$ where $s_k$ refers to sample $k$ for $k=1,2,...,K$;  $y_i$ is the response of unit $i$; $w_{ij}$ is the weight of the link from unit $i$ to unit $j$; $w_i^+$ is the sum of the weights of all links emanating from individual $i$; $t_{k,i}$ is the step in the sampling sequence that unit $i$ is selected for sample $k$; $\underline{J}_k$ and $\underline{H}_k$ are indicator vectors of length $L=\underset{k=1,2,...,K}{\text{max}}\{n_{k}\}$ that record when random jumps are taken and when the active set is exhausted in the sample selection procedure so that a random jump is forced at this step, respectively. Note that $H_{k,t}=1$ implies $J_{k,t}=1$ but the converse will only always hold when $d=1$. It shall be understood that for all $k=1,2,...,K$, $J_{k,1},...,J_{k,n_{0k}}=H_{k,1},...,H_{k,n_{0k}}=0$ and if $n_k<L$ then $J_{k,n_k+1},...,J_{k,L}=H_{k,n_k+1},...,H_{k,L}=0$.

\begin{figure}[H]
\begin{center}
\includegraphics [width=2.6 in,height=1.2 in]{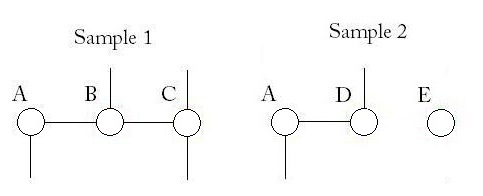}
\end{center}
\vspace{-5mm}
\caption{A two-sample study where samples are selected via the design that permits for random jumps. The out-degree of each node is equal to the number of links emanating from the node.}
\label{jumpsexampleappendix}
\end{figure}

We clarify the notation and sample selection procedure with the following example. Figure \ref{jumpsexampleappendix} provides an example of two final samples selected under a design that permits for random jumps where the study is comprised of two samples, thus $K=2$. The size of the initial random samples are $n_{01}=n_{02}=1$ and the number of members added after the initial samples is two, to bring the final sample sizes up to $n_1=n_2=3$. In each case the active set is always the current sample.
\vspace{-5mm}

Suppose that links between nodes are reciprocated and the weight of each link is set equal to one. Further suppose that $s_{(0_1,0_2)}=((A,B,C),(E,D,A))$.

First consider that $0<d<1$. For sample 1, suppose unit $B$ is added via tracing a link and unit $C$ is added via a random jump. For sample 2, suppose unit $D$ is added via taking a random jump (which must be forced at this step) and unit $A$ is added via tracing a link. Then $\underline{J}_1\equiv \underline{J}_1^{(0_1)}=(0,0,1),\underline{H}_1\equiv\underline{H}_1^{(0_1)}=(0,0,0),\underline{J}_2\equiv \underline{J}_2^{(0_2)}=(0,1,0)$, and $\underline{H}_2\equiv\underline{H}_2^{(0_2)}=(0,1,0)$ are the original $\underline{J}$ and $\underline{H}$ vectors and hence $\underline{\mathcal{J}}\equiv \mathcal{J}^{(0_1,0_2)}=(0,1,1)$. Permitting a slight abuse of notation, we leave it implicit within the probability expressions that the corresponding adjacency data is observed. Now, the probability of obtaining the original data is
\begin{align}
p(s_{(0_1,0_2)},\underline{J}_1^{(0_1)},\underline{H}_1^{(0_1)},\underline{J}_2^{(0_2)},\underline{H}_2^{(0_2)})=\bigg(\frac{1}{N}(d)\frac{1}{2}(1-d)\frac{1}{N-2}\bigg)\times\bigg(\frac{1}{N}\frac{1}{N-1}(d)\frac{1}{2}\bigg).
\end{align}

If $d=1$, so that random jumps are permitted only when links out of the active set are exhausted, and $s_{(0_1,0_2)}=((A,B,C),(E,D,A))$ then $\underline{J}_1\equiv \underline{J}_1^{(0_1)}=(0,0,0),\underline{H}_1\equiv\underline{H}_1^{(0_1)}=(0,0,0),\underline{J}_2\equiv \underline{J}_2^{(0_2)}=(0,1,0)$, and $\underline{H}_2\equiv\underline{H}_2^{(0_2)}=(0,1,0)$ must be the original $\underline{J}$ and $\underline{H}$ vectors and hence $\underline{\mathcal{J}}\equiv \mathcal{J}^{(0_1,0_2)}=(0,1,0)$. The probability of obtaining the original data is
\begin{align}
p(s_{0_1,0_2})=\bigg(\frac{1}{N}\frac{1}{2}\frac{1}{3}\bigg)\times\bigg(\frac{1}{N}\frac{1}{N-1}\frac{1}{2}\bigg).
\end{align}

\subsection{Sufficiency result}

The reduced data is defined as $r_d(d_0)=d_r=\{(i,y_i,w_{ij},w_i+),\underline{\mathcal{J}}:i,j\ \epsilon\ s_k,\ k=1,2,...,K\}$, where $r_d$ is the reduction function and $\underline{\mathcal{J}}=(\sum\limits_{k=1}^K J_{k,1},\sum\limits_{k=1}^K J_{k,2},...,\sum\limits_{k=1}^K J_{k,L})
=(\mathcal{J}_1,\mathcal{J}_2,...,\mathcal{J}_L)$. In this case the reduction function removes the time/step element assigned to each unit selected for each sample, removes the $\underline{H}_k$ vectors, and reduces the records of when random jumps are taken to a sum of the number of random jumps taken at the corresponding steps across all samples.

We define $\underline{\theta}=(N,\underline{y}_N,\underline{w}_N,\underline{w}_N^+)$ to be the parameter of interest where $N$ is the population size, $\underline{y}_N$ is the vector of length N which displays the individual responses, $\underline{w}_N$ is the adjacency matrix of size $N \times N$ of the population graph, and $\underline{w}_N^+$ is a vector of length $N$ which displays the out-degree of the members of the population. We make the definition that $\underline{\theta}$ is $\textit{consistent}$ with the reduced data $d_r$ if $\underline{\theta}$ can be arranged such that the first $n=|\bigcup\limits_{k=1}^K s_{k}|$ units share the same pattern of selection over the $K$ samples as well as the structure in terms of responses of interest, links within the samples, and out-degree, as those in the final samples. Alternatively, we can say the corresponding reduced data of these elements is equivalent to $d_r$. For notational convenience we refer to the ordered sets of these responses as $\underline{y}_{d_r}$, $\underline{w}_{d_r}$ and $\underline{w}_{d_r}^+$, respectively. Keeping consistent with the theoretical setup for a design-based approach in the usual survey sampling setting (see \cite{Thompson1996} for more information) the set of all $\underline{\theta}$ consistent with the reduced data $d_r$ shall be labeled as $\Theta_{d_r}$. Notice that since the population size is unknown, $N$, and hence the lengths of vectors and sizes of matrices corresponding with $\underline{y}_N$, $\underline{w}_N$ and $\underline{w}_N^+$, is permitted to range over all values in the natural number set.

\bigskip

\noindent \textbf{Theorem}: When samples are obtained with the sampling design that permits for random jumps, $D_r$ as defined above is a sufficient statistic for $\underline{\theta}=(N,\underline{y}_N,\underline{w}_N,\underline{w}_N^+)$.

\bigskip

\noindent \textbf{Proof}: First consider sample $k$ and step $t=1,2,...,n_k-n_{0k}$. Recall that $J_{k,t+n_{0k}}=0$ if a link is traced and $J_{k,t+n_{0k}}=1$ if a random jump is taken at step $t$ of the selection of sample $k$. Also recall that $H_{k,t+n_{0k}}=1$ if $w_{a_{k,t},+}=0$, that is, a random jump is forced at this step in the sample selection procedure as there are no links to trace out at step $t$, and 0 otherwise. For $t>0$ we define $q_{k,t,i}$ to be the probability of obtaining that unit selected at step $t$ for sample $k$ if the unit is added via tracing a link; otherwise we take $q_{k,t,i}=1$. For $t=0$ we take $q_{k,t,i}=\frac{1}{d}$.

Now, let $d_0$ be any data point where $P(D_0=d_0)>0$. Then,
\begin{align}
P_{\underline{\theta}}(D_0=d_0)&=P(D_0=d_0\mid N,\underline{y}_N,\underline{w}_N,\underline{w}_N^+)I[\underline{\theta}\epsilon \Theta _{d_r}]\notag \\
&=P(D_0=d_0\mid N,\underline{y}_{d_r},\underline{w}_{d_r},\underline{w}_{d_r}^+)I[\underline{\theta}\epsilon \Theta _{d_r}]\notag \\
&=\prod\limits_{k=1}^K \bigg\{\frac{1}{{N \choose n_{0k}}}
\prod\limits_{t=0}^{n_k-n_{0k}}\bigg[d q_{k,t,i}^{(1-J_{k,t+n_{0k}})}\times\notag\\
&\bigg((1-d)\frac{1}{N-(n_{0k}+t-1)}\bigg)^{J_{k,t+n_{0k}}(1-H_{k,t+n_{0k}})} \times \notag \\
&\bigg(\frac{1}{N-(n_{0k}+t-1)}\bigg)^{J_{k,t+n_{0k}}(H_{k,t+n_{0k}})}\bigg]\bigg\}
I[\underline{\theta}\epsilon \Theta_{d_r}]\notag\\
&=\prod\limits_{k=1}^K \bigg\{\bigg(\prod\limits_{t=0}^{n_k-n_{0k}}(d q_{k,t,i})^{(1-J_{k,t+n_{0k}})}\bigg)
(1-d)^{\sum\limits_{t=0}^{n_k-n_{0k}}J_{k,t+n_{0k}}(1-H_{k,t+n_{0k}})}\bigg\}\times \notag \\
&\prod\limits_{k=1}^K \frac{1}{{N \choose n_{0k}}} \prod\limits_{i=1}^L\bigg(\frac{1}{N-(i-1)}\bigg)^{\mathcal{J}_i}
I[\underline{\theta}\epsilon \Theta_{d_r}] \notag\\
&=h(d_0)\times g(d_r,\underline{\theta}).
\label{likelihoodjumps}
\end{align}

\noindent Therefore, by the Fisher-Neyman Factorization Theorem, $D_r$ is a sufficient statistic for $\underline{\theta}=(N,\underline{y}_N,\underline{w}_N,\underline{w}_N^+)$.

$\Box$

\subsection{Consistent data reorderings; discussion and examples}

Referring back to the example presented in Figure \ref{jumpsexampleappendix}, if $0 < d < 1$ then one pair of sample reorderings consistent with the sufficient statistic is $s_{(x_1,x_2)}=((C,A,B),(D,A,E))$ if we allow for a random jump to be taken when unit $A$ is added to the corresponding reordering for sample 1. Notice that this requires unit $A$ to be added via tracing a link from unit $D$ in sample 2, since there is only one jump taken at this point in the combined sample selection procedure. In this case, $\underline{J}_1^{(x_1)}=(0,1,0), \underline{H}_1^{(x_1)}=(0,0,0)$ and $\underline{J}_2^{(x_2)}=(0,0,1),\underline{H}_2^{(x_2)}=(0,0,0)$ so that $\underline{\mathcal{J}}^{(x_1,x_2)}=(0,1,1)$ is consistent with $\underline{\mathcal{J}}$. The probability of obtaining this pair of reorderings is then
\begin{align}
p(s_{(x_1,x_2)},\underline{J}_1^{(x_1)},\underline{H}_1^{(x_1)},\underline{J}_2^{(x_2)},\underline{H}_2^{(x_2)})
=\bigg(\frac{1}{N}(1-d)\frac{1}{N-1}(d)\frac{2}{5}\bigg)\times\bigg(\frac{1}{N}(d)\frac{1}{2}(1-d)\frac{1}{N-2}\bigg).
\end{align}
However, the pair of reorderings $s_{(x_1,x_2)}=((C,A,B),(E,A,D))$ is not consistent with the sufficient statistic as it requires a random jump to be made at the second step of selection for both samples. This will result in the second entry of $\underline{\mathcal{J}}^{(x_1,x_2)}$ to be equal to 2 and hence will not be consistent with $\underline{\mathcal{J}}$. Therefore this pair of sample reorderings is not consistent with the sufficient statistic.

If $d=1$ then $s_{(x_1,x_2)}=((A,B,C),(E,A,D))$ is a pair of reorderings consistent with the sufficient statistic, and which turns out to share the same $\underline{J}$ and $\underline{H}$ vectors as the original sample reorderings. This pair has empirical probability, that is, the probability of obtaining this reordering in the full population graph setting, of selection
\begin{align}
p(s_{x_1,x_2})=\bigg(\frac{1}{N}\frac{1}{2}\frac{1}{3}\bigg)\times\bigg(\frac{1}{N}\frac{1}{N-1}\frac{1}{2}\bigg).
\end{align}
However, $s_{(x_1,x_2)}=((C,A,B),(E,A,D))$ is a pair of sample reorderings that are not consistent with the sufficient statistic, since it has zero probability of being selected due to an absence of a link to trace from unit $C$ to unit $A$ in the first sample.

\end{document}